\documentclass[preprint,showpacs,preprintnumbers,amsmath,amssymb]{revtex4}

\usepackage{graphicx}% Include figure files
\usepackage{dcolumn}% Align table columns on decimal point
\usepackage{bm}% bold math

\newcommand{\be}{\begin{equation}}  
\newcommand{\ee}{\end{equation}}  
\newcommand{\bear}{\begin{eqnarray}}  
\newcommand{\eear}{\end{eqnarray}}  
\newcommand{\ba}{\begin{array}}  
\newcommand{\ea}{\end{array}}

\newskip\humongous \humongous=0pt plus 1000pt minus 1000pt

\newif\ifdtup

\def\oldreffmt#1{\rlap{[#1]} \hbox to 2\parindent{}}

\def\figfmt#1{\rlap{Figure {#1}} \hbox to 1in{}}  
  
\def\ie{\hbox{\it i.e.}{}}	  
\def\eg{\hbox{\it e.g.}{}}

\def\bra#1{\left\langle #1\right|}  
\def\ket#1{\left| #1\right\rangle}  
\def\VEV#1{\left\langle #1\right\rangle}

\def\slash#1{#1\!\!\!/\!\,\,}  
\def\beq{\begin{equation}}  
\def\eeq{\end{equation}}  
\def\bea{\begin{eqnarray}}  
\def\eea{\end{eqnarray}}  
\def\half{\frac{1}{2}}  
  
\def\bq{\begin{quote}}  
\def\eq{\end{quote}}

\def\half{\frac{1}{2}}       
\def \lta {\mathrel{\vcenter  
     {\hbox{$<$}\nointerlineskip\hbox{$\sim$}}}}  
\def \gta {\mathrel{\vcenter  
     {\hbox{$>$}\nointerlineskip\hbox{$\sim$}}}}   
\relax  

\newdimen\tdim  
\tdim=\unitlength  
\def\bar{\overline}

\begin{document}

\preprint{FERMILAB-PUB-12-589-T}

\vspace{2.0cm}

\title{``Super''--Dilatation Symmetry of the Top-Higgs System}

\author{Christopher T. Hill}

\email{hill@fnal.gov}

\affiliation{
 {{Fermi National Accelerator Laboratory}}\\
{{\it P.O. Box 500, Batavia, Illinois 60510, USA}}
}%

\date{\today}% It is always \today, today,
             %  but any date may be explicitly specified

\begin{abstract}
The top-Higgs system, consisting of top quark ({\em LH} doublet, {\em RH} singlet) 
and Higgs boson kinetic
terms, with gauge fields set to zero, has an {\em exact} (modulo
total divergences) symmetry where 
both fermion and Higgs fields are 
shifted and mixed in a supersymmetric fashion.
The full Higgs-Yukawa interaction and Higgs-potential, including
additional  $\sim 1/\Lambda^2$ NJL-like interactions, also has this
symmetry to ${\cal{O}}(1/\Lambda^4)$, up to null-operators. 
Thus the interaction lagrangian can be viewed
as a power series in $1/\Lambda^2$.
The symmetry involves interplay
of the Higgs quartic interaction with the Higgs-Yukawa interaction
and implies the relationship, $\lambda = \half g^2$ between the
top--Yukawa coupling, $g$, and Higgs quartic coupling,
$\lambda$, at a high energy scale $ \Lambda \gta$ few TeV. 
We interpret this to be a new physics scale. The top
quark is massless in the symmetric phase, satisfying
the Nambu-Goldstone theorem.
The fermionic shift part of the current is $\propto (1-H^\dagger H/v^2)$,
owing to the interplay of $\lambda$ and $g$,
and vanishes in the broken phase. Hence the Nambu-Goldstone
theorem is trivially evaded in the broken phase 
and the top quark becomes heavy (it is not a Goldstino).
We have $m_t=m_h$,
subject to radiative corrections that can in principle
pull the Higgs into concordance with experiment. 
\end{abstract}

% PACS, the Physics and Astronomy
                             % Classification Scheme.
%\keywords{Suggested keywords}%Use showkeys class option if keyword
                              %display desired
\maketitle

\noindent
\section{Introduction}
\vskip .1in
\noindent

One of the remarkable features of the newly discovered Higgs boson is
the proximity of three of the defining quantities of the standard model:
\beq
\label{oneone}
 m^2_{Higgs} \approx \half m_{top}^2 \qquad m_{top} \approx v_{weak} 
\eeq
where $v_{weak}\approx 175$ GeV is the Higgs VEV 
(Note: in the following $m_h$ will denote the physical particle mass of the Higgs boson, while $M_H^2H^\dagger H$ will be the lagrangian mass term)
The part of the standard model lagrangian that subsumes 
these relationships is the top-Higgs system:
\beq
{\cal{L}} = {\cal{L}}_{kinetic}+ g\bar{\psi}_Lt_RH + h.c. -\frac{\lambda}{2}\left( H^\dagger H - v_{weak}^2\right)^2
\eeq
where $\psi = (t, b)$ 
\footnote{For notational reference: The physical real Higgs boson field $h$, defined by $v$ at the minimum of the Higgs potential, is  given by $H=v+h/\sqrt{2}$, with mass $m_h^2 = 2\lambda v^2$.
When the potential is written in the form $-M_H^2 H^\dagger H +(\lambda/2)(H^\dagger H)^2$ we have the correspondence $M_H^2 = \lambda v^2
=m_h^2/2$}.
Eq.(\ref{oneone}) corresponds to the following choice of top Higgs-Yukawa, 
and quartic coupling constants at the scale of the top quark and Higgs boson masses:
\beq
g \approx 1, \qquad\qquad   {\lambda} \approx \frac{1}{4}
\eeq
This is astonishingly consistent with experiment to within a few percent.
These relationships are suggestive of some
hidden dynamical or symmetrical interplay 
between the top quark and Higgs boson.

Our main interest is in mechanisms that can relate
the quantities
$m_h$, $m_t$ and $v_{weak}$.
 One idea is that of a composite Higgs boson, 
\eg, composed of $\bar{t}t$. This is an old approach 
\cite{yamawaki,BHL}, typically treated in the context of the
Nambu-Jona-Lasinio model, which implies some new underlying 
strong dynamics, such as topcolor \cite{topc}. The pure fermion
loop approximation NJL model yields $m_h = 2 m_t$, while fine-tuning
a large hierarchy between the Higgs composite scale
and the weak scale  leads to a lighter Higgs
boson, but $m_{h} \sim 260$ GeV, and $m_{t}\sim 220$ GeV, given the input
$v_{weak} = 175$ GeV. Still other approaches avoid the fine-tuning, yielding
$m_h\sim 1$ TeV.
Hence $\bar{t}t$ composite Higgs models do succeed in relating
$m_h$, $m_t$ and $v_{weak}$, but generally predict a Higgs boson
that is significantly heavier than the top quark \cite{BHL}.  We would have expected a light pNGB,
such as the top-pion \cite{topc}, in lieu of a light Higgs boson with
the tree-level rates for $h\rightarrow ZZ^*$ (a pNGB would have $\pi \rightarrow Z Z^*$
at the significantly lower, axial anomaly level, rather than at the observed tree-level). 

In the present paper we examine an alternative idea: we propose
a generalization of dilatation symmetry for the Higgs boson
that involves a ``super"-symmetric 
relationship between the top and Higgs fields.  
Indeed, this symmetry is exact in the 
top, with bottom-left, and Higgs kinetic terms, in the gaugeless limit
(up to total divergences; it does not
rely upon equations of motion).

This symmetry is similar to an ordinary c-number dilatation in the sense that
it shifts the Higgs boson field, but the shift of the Higgs is now promoted 
to an operator. Our transformation also shifts the fermions. Finally,
it includes a supersymmetric rotation, or ``twist,''  between the fermions 
and the Higgs field. 
The twist and shifts are locked together by the invariance
of the Higgs mass term,
the Higgs quartic interaction and the Higgs-Yukawa coupling constant
in a nontrivial way. While the symmetry is exact in
the (gaugeless) kinetic terms, we do make use of the fermion
equations of motion (\ie, we set null-operators to zero)
in the  transformation of the Higgs-Yukawa interaction.
This has the nontrivial consequence of modifying
the current in such a a way that the fermionic shifts are present
in the unbroken phase of the theory, but ``turn off'' in the broken phase.

Indeed, the symmetry yields
$\lambda = g^2/2$ in both phases of the Standard
Model.  However, the fermionic shift part of the current
is nontrivially modified by the quartic-Yukawa interplay
and is $\propto (1-H^\dagger H/v^2)$. It thus vanishes in
the broken phase with $\VEV{H}=v$.   Therefore,
the top quark can become massive in the broken phase in the usual way, with
the relationship $m_h = m_t$. This is consistent with the Nambu-Goldstone theorem that would
otherwise naively force the top quark to be 
a massless Goldstino.   
This relationship $\lambda = g^2/2$ 
is the analogue of a Goldberger-Treiman'' relationship.
It holds at a high scale, $\Lambda$, and
is subject to renormalization group and higher dimension operator effects that can bring the physical masses into concordance with $m_h^2 \approx m_t^2/2$.

While the usual superalgebra of SUSY does not permit
the Higgs to be the superpartner of the $(t,b)$ quarks,
the symmetry we present here does accomplish this.
We emphasize at the outset that the ``super''-dilatation symmetry is not
a conventional superalgebra, \ie, it is not a grading of the Lorentz Group, and is not associated with a nontrivial nonabelian closed superalgebra
(at least not in our present exploratory formulation). 
The symmetry is a bosonic $U(1)$ invariance, and therefore closes 
trivially.  As such, there is some {\em a priori} freedom in the normalization of
its generators. The freedom is partially fixed by the exact invariance
of the top and Higgs kinetic terms
in the limit that gauge bosons are switched off (\ie, global
gauge symmetries remain). It is completely fixed by super-mixing between
Higgs mass, quartic and Yukawa interactions.

Formally, this is 
remniscent of a nonlinearly realized supersymmetry
but it isn't: it's actually linearly realized.  Operationally it is also
similar to a ``reparameterization invariance," \eg, as occurs
in heavy quark effective field theory (HQET) \cite{Georgi,LukeManohar,Heinonen}.  
In the latter case 
one considers an $M\rightarrow \infty$ limit for a heavy quark and constructs a  field theoretic lagrangian 
for a given four-velocity ``supersector,'' $v_\mu$. The lagrangian
takes the form of a series expansion in higher dimension operators weighted by powers of $1/M$.  The leading terms in the theory
display heavy-spin symmetry (\eg, degenerate  $0^-$ and $1^{-}$ mesons). 
The reparameterization invariance is a residual symmetry that constrains 
the full operator structure and
relates the coefficients of the terms in the 
lagrangian to higher orders of $1/M$.  The
reparameterization invariance is essentially the vestige
of the underlying hidden Lorentz invariance \cite{Heinonen}.

We will presently move directly to discuss our symmetry. 
However, in Appendix A we review the analogous dilatonic Higgs symmetry
to remind  the reader of the interplay of a ``shift'' symmetry with a 
dilatation.  The dilaton is inherited from a shift invariance,
\eg, the ``modular symmetry'' of a flat potential, which is then
incorporated into the scale transformation. We show
explicitly how this works, because we feel it is
a confusing point in these schemes.  The dilaton current
comes from the shift-current and {\em  not} the scale current,
which is $S_\mu \sim x^\nu T_{\mu\nu}$. Here $T_{\mu\nu}$ is the 
``improved stress tensor'' and involves only second derivatives of bosons
whereas the dilatonic current is first order in derivatives.
In Higgs-as-dilaton schemes we can always start in
the unbroken phase, and the orientation of the Higgs boson in group space must be specified when a dilatonic shift is performed. This is in analogy
to the fermionic orientations we encounter here. The reader may wish
to consult Appendix A to see the analogy with our present construction.

In Appendix B we give, for motivation sake, the analogy
of our present approach to a ``bottoms-up'' derivation of
a nonlinear chiral lagrangian and the Goldberger-Treiman relation.  This is completely parallel to the present
derivation. We will initially consider the top and Higgs
kinetic terms of the standard model with gauge fields set to zero:
\beq
\label{kt0}
 {\cal{L}}_{K} = \bar{\psi}_L i\slash{\partial}\psi_L + \bar{t}_R i\slash{\partial} t_R +\partial H^\dagger \partial H 
\eeq
These kinetic term ${\cal{L}}_K$ will be exactly
invariant under our shift symmetry, modulo total divergences. We will then 
consider the potential terms: 
\bea
\label{interaction}
{\cal{L}}_V & = &  
 g(\bar{\psi}_L t_R H + h.c.)
- M_H^2 H^\dagger H - \frac{\lambda}{2} (H^\dagger H)^2
\eea
 We will
demand $ {\cal{L}}_V $ be invariant under the symmetry.  
We will be able to impose the symmetry, up to null operators
(operators that vanish by equations of motion). This will
yield the analogue of the ``Goldberger-Treiman'' relation,
$\lambda = g^2/2$. It will also require the introduction
of higher dimension operators $\sum_n {\cal{O}}^n\propto 1/\Lambda^{2n}$.  
At present we don't know how to sum this series, and we
will truncate at the scale $1/\Lambda^4$. Hence, to order
$1/\Lambda^4$ we'll have a lagrangian that is fully invariant under
our symmetry.  We will refer to our
main relationship, $\lambda =g^2/2$, or $m_h =\sqrt{2}|M_H| = m_t$ as
the analogue of a
``Goldberger-Trieman'' relation  in this context.

\section{``Super"--Shift Symmetry of the Top-Higgs Kinetic Terms}
 
We begin by introducing fixed orientations, $\theta_{L,R}$. 
These will be treated, with scale factors, as shifts
of the quark fields themselves, $\delta\psi_L= \theta_L\epsilon +...$
and $\delta t_R = \theta_R\epsilon +...$, where $\psi_L = (t, b)_L$.  
The $\theta_{L,R}$ can be viewed as fixed zero 4-momentum components of the top (and bottom) fermion fields that have an arbitrary alignment in spin-space,
and gauge group spaces of color and electroweak interactions. 
The $\theta_{L,R}$ are assumed to carry 
the same global color, isospin, weak hypercharge
and baryon number quantum numbers as the
corresponding $\psi_L$ and $t_R$ fermion fields, ie, $\theta_L \sim (N_c=3,I=\half, Y=\frac{1}{3},B=\frac{1}{3})$ and $\theta_R \sim (N_c=3,I=0, Y=\frac{4}{3},B=\frac{1}{3})$ . 
In a sense, one could view the $\theta_{L,R}$
as constant VEV's of the corresponding fermion fields,
but they have no physical effect since the theory will be
assumed symmetric under these small shifts.
Our transformation
is done for arbitrary fixed values of the $\theta_{L,R}$ The
parameter is an 
infinitesimal bosonic c-number,  $\epsilon$.
  The  $\theta_{L,R}$ are considered to
have scale dimension-$\frac{3}{2}$, (in analogy to the dilatonic Higgs case of eq.(\ref{five20}) where we have scale-dimension $1$). We thus introduce a scale $\Lambda$ into our transformation,
that is to be viewed as ``large" compared to the weak scale.

Consider the top and Higgs
kinetic terms of the standard model with gauge fields set to zero:
\beq
\label{kt}
 {\cal{L}}_{K} = \bar{\psi}_L i\slash{\partial}\psi_L + \bar{t}_R i\slash{\partial} t_R +\partial H^\dagger \partial H 
\eeq
We define the infinitesimal transformation:
 \bea
\label{trans0}
\delta \psi^{ia}_L = \theta^{ia}_L\eta \epsilon - i\frac{\slash{\partial}H^i\theta^a_R}{\Lambda^2}\epsilon ; &&
\qquad  
\delta \bar{\psi}_{L\;ia} = \bar{\theta}_{L\;ia}\eta \epsilon + 
i\frac{\bar{\theta}_{Ra}\slash{\partial}H_i^\dagger}{\Lambda^2}\epsilon  ;
\nonumber  \\
\delta t^a_R = \theta^a_R\eta \epsilon 
-i \frac{\slash{\partial}H_i^\dagger\theta^{ia}_L}{\Lambda^2}\epsilon ;  &&
\qquad  
\delta \bar{t}_{Ra} = \bar{\theta}_{Ra}\eta \epsilon 
+ i\frac{\bar{\theta}_{Lia}\slash{\partial}H^i}{\Lambda^2}\epsilon  ;
\nonumber  \\
\delta H^i = \frac{\bar{\theta}_{Ra}\psi^{ia}_L + \bar{t}_{Ra}\theta^{ia}_L}{\Lambda^2}\epsilon ; 
&&
\qquad  \delta H_i^\dagger = \frac{\bar{\psi}_{Lai}\theta^a_R + \bar{\theta}_{Lai}t^a_R}{\Lambda^2}\epsilon .
\eea
where $i$ ($a$) is an 
isospin (color) index. 
$\eta$ is a relative normalization factor that we determine
subsequently.
Presently we are not including a c-number shift in $H$,
(as in eq.(\ref{five20}) of Appendix A),
into the full transformation.  The above
transformation is our most general case, but
it is convenient to consider more minimal 
versions of the symmetry
where either $\theta_L$ or $\theta_R$ are set to zero 
or equal (see section III).

It is readily seen that eq.(\ref{trans0}) is an invariance of eq.(\ref{kt})
up to total derivatives:
\bea
\label{transkt}
\delta (\bar{\psi}_Li\slash{\partial}\psi_L) & = & \frac{(\bar{\psi}_L\theta_R)\cdot \partial^2 H}{\Lambda^2}\epsilon + h.c.+t.d.
\nonumber  \\
\delta (\bar{t}_Ri\slash{\partial}t_R) & = & \frac{ (\bar{\theta}_L t_R)\cdot \partial^2 H}{\Lambda^2}\epsilon + h.c.+t.d.
\nonumber  \\
\delta (\partial H^\dagger \partial H) & = & -\frac{(\bar{\psi}_L\theta_R + \bar{\theta}_Lt_R)\cdot\partial^2H}{\Lambda^2}\epsilon
+ h.c. + t.d.
\nonumber \\
\makebox{hence,} & &  \delta  {\cal{L}}_{K}  =  0 + t.d.
\eea
(we've suppressed indices; ``t.d.'' refers to ``total divergence''). 
The symmetry of the gauge free kinetic terms 
makes no use of equations of motion or on-shell conditions.
At this stage,  the shifts 
in $\psi_L$ and $t_R$ by $\theta_{L,R}$ proportional to $\eta$
play no role, but
will be essential with the Higgs-Yukawa interaction and Higgs mass term. Indeed, shifting $\bar{\psi}_Li\slash{\partial}\psi_L
\rightarrow \bar{\psi}_Li\slash{\partial}\theta_L$ yields a total divergence
provided we have switched off the local gauge fields.

This transformation exploits the interplay of the quantum 
numbers of $\psi_L$, $t_R$ and $H$. It resembles a scalar supermultiplet 
transformation of component fields \cite{Wess}, where 
the Higgs field is treated as a superpartner
of $\psi_L$.  We emphasize that this is not a representation of the supersymmetry algebra (there is no ``F'' auxillary field, \cite{Wess}; 
this is essentially  a scalar supermultiplet transformation 
with fixed $F=0$ and the superparameters replaced by $\theta\epsilon$).
With the assignment of scales of the $\theta_{L,R}$
and the presence of $\Lambda$ the commutators of subsequent transformations for
different $\theta_{L,R}$ cannot close. Also, the $\theta_{L,R}$ carry
flavor and color quantum numbers, and the failure of the algebra to close
into a superalgebra is presumably a supersymmetric extension of the Coleman-Mandula no-go theorem.   In fact, this is a $U(1)$ symmetry with the transformation parameter, $\epsilon$, 
for fixed background values of $\theta_{L,R}$. As such, the commutator 
trivially vanishes on the fields:
 \beq
 [\delta_{\epsilon{}'}, \delta_{\epsilon}](\psi, H, t_R) = 0
 \eeq

We presently turn to the full lagrangian
of the top-Higgs system in the standard model with gauge fields turned off:
\bea
\label{full}
{\cal{L}}_H & = &  i\bar{\psi}_L \slash{\partial}\psi_L + i\bar{t}_R \slash{\partial} t_R +\partial H^\dagger \partial H 
\nonumber \\
& & + g(\bar{\psi}_L t_R H + h.c.)
- M_H^2 H^\dagger H - \frac{\lambda}{2} (H^\dagger H)^2
\eea
From eq.(\ref{trans0}) we compute the transformations:
\bea
\label{trans2}
\delta (-M_H^2 H^\dagger H) & = & 
-\frac{\epsilon}{\Lambda^2} \; M_H^2(\bar{\psi}_{L}\theta_R + \bar{\theta}_{L}t_R)\cdot H + h.c
 \\
 \label{transquartic}
\delta (-\frac{\lambda}{2}(H^\dagger H)^2) & = & -\frac{\epsilon}{\Lambda^2}\; \lambda(\bar{\psi}_{L}\theta_R + \bar{\theta}_{L}t_R)\cdot H H^\dagger H  + h.c.
\\
\label{transYuk}
\delta (g\bar{\psi}_L t_R H + h.c.) & = & 
g\eta\epsilon(\bar{\psi}_L\theta_R  
+ \bar{\theta}_L t_R) H  
+g^2\frac{\epsilon}{2\Lambda^2} (\bar{\theta}_R \psi_L
+ \bar{t}_{R} \theta_L)\cdot \left( H^\dagger H^\dagger H \right) 
 \nonumber \\
& & + \; g\frac{\epsilon}{\Lambda^2}\bar{\psi}_L t_R(\bar{\theta}_{R}\psi_L + \bar{t}_{R}\theta_L)
\nonumber \\
& & +ig\frac{2\epsilon}{\Lambda^2}\bar{\psi}_L\gamma_\mu\frac{\tau^A}{2}\theta_L 
\left( H^\dagger \stackrel{\leftrightarrow}{\partial^\mu} \frac{\tau^A}{2} H \right)
+ig\frac{\epsilon}{2\Lambda^2} \bar{\psi}_L\gamma_\mu\theta_L 
\left( H^\dagger \stackrel{\leftrightarrow}{\partial^\mu} H \right)
\nonumber \\
& &
-ig\frac{\epsilon}{\Lambda^2}\bar{\theta}_{R}\gamma_\mu t_R 
\left( H^\dagger \stackrel{\leftrightarrow}{\partial^\mu} H \right)  + h.c. + t.d.
\eea
Eq.(\ref{transYuk}) is crucial to our construction.
It is obtained as follows:
\bea
\label{transYuk2}
\delta (g\bar{\psi}_L t_R H + h.c.) & = &
g\left(\bar{\psi}_L (\theta_R\eta \epsilon -i\epsilon\frac{\slash{\partial}H^\dagger \theta_L}{\Lambda^2} ) H 
 +  
(\bar{\theta}_L\eta \epsilon + i\epsilon 
\frac{\bar{\theta}_{R}\slash{\partial}H^\dagger}{\Lambda^2} )t_R H \right) 
\nonumber \\
& & + \; g\epsilon\left(\bar{\psi}_L t_R(\bar{\theta}_{R}\psi_L  +  \bar{t}_{R}\theta_L)\frac{1}{\Lambda^2}\right)
+ h.c. 
\nonumber \\
&  = & g\eta\epsilon(\bar{\psi}_L\theta_R  
+ \bar{\theta}_L t_R) H  
 + \; \frac{g\epsilon}{\Lambda^2}\bar{\psi}_L t_R(\bar{\theta}_{R}\psi_L  +  \bar{t}_{R}\theta_L)
\nonumber \\
& & +\; i\frac{g\epsilon}{2\Lambda^2} (\partial^\mu \bar{\psi}_L \gamma_\mu )\cdot H  (H^\dagger \cdot \theta_L) 
-i \frac{g\epsilon}{2\Lambda^2}\bar{\theta}_{R}(\gamma_\mu \partial^\mu t_R) (H^\dagger \cdot H)
\nonumber \\
& & +\; i\frac{2g\epsilon}{\Lambda^2}\bar{\psi}_L\gamma_\mu\frac{\tau^A}{2}\theta_L 
\left( H^\dagger \stackrel{\leftrightarrow}{\partial^\mu} \frac{\tau^A}{2} H \right)
 +i\frac{g\epsilon}{2\Lambda^2}\bar{\psi}_L\gamma_\mu\theta_L 
 \left( H^\dagger \stackrel{\leftrightarrow}{\partial^\mu} H \right)
\nonumber \\
& &
-\; i\frac{g\epsilon}{\Lambda^2} \bar{\theta}_{R}\gamma_\mu t_R 
\left( H^\dagger \stackrel{\leftrightarrow}{\partial^\mu} H\right)  + h.c. + t.d.
\eea
where we use the isospin Fierz identity,  $[\tau^A]_{ij} [\tau^A]_{kl} = 2\delta_{il}\delta_{kj} -\delta_{ij}\delta_{kl}$, 
and: $\stackrel{\leftrightarrow}{\partial^\mu}=\half(\stackrel{\rightarrow}{\partial^\mu}
-\stackrel{\leftarrow}{\partial^\mu})$.
We then apply the fermionic equations of motion to eq.(\ref{transYuk2}):
\beq
i\slash{\partial}t_R + g\psi_L\cdot H^\dagger = 0
\qquad \qquad
i\slash{\partial} {\psi}_L + gt_R H =0
\eeq 
and eq.(\ref{transYuk}) follows.

Notice in eq.(\ref{transYuk}) we have 
generated higher dimension operator
terms of the form:
\bea
& & \frac{g\epsilon}{\Lambda^2}\bar{\psi}_L t_R(\bar{\theta}_{R}\psi_L  +  \bar{t}_{R}\theta_L)
 +\; i\frac{2g\epsilon}{\Lambda^2}\bar{\psi}_L\gamma_\mu\frac{\tau^A}{2}\theta_L 
 (H^\dagger \stackrel{\leftrightarrow}{\partial^\mu} \frac{\tau^A}{2} H )
\nonumber \\
& &
 +i\frac{g\epsilon}{2\Lambda^2}\bar{\psi}_L\gamma_\mu\theta_L 
 ( H^\dagger \stackrel{\leftrightarrow}{\partial^\mu} H )
-\; i\frac{g\epsilon}{\Lambda^2} \bar{\theta}_{R}\gamma_\mu t_R 
( H^\dagger \stackrel{\leftrightarrow}{\partial^\mu} H )  + h.c. + t.d.
\eea
In analogy to the ``bottoms up'' derivation
of a nonlinear chiral lagrangian (see Appendix B), 
these terms can be cancelled by adding
higher dimension operators to the original
lagrangian of the form:
\bea
\label{D6}
 {\cal{L}}_{d=6} & = &  \frac{\kappa}{\Lambda^2}(\bar{\psi}_L t_R\bar{t}_{R}\psi_L )
+\frac{2\kappa}{\Lambda^2}(\bar{\psi}_L\gamma_\mu\frac{\tau^A}{2}\psi_L )
(H^\dagger i\stackrel{\leftrightarrow}{\partial^\mu} \frac{\tau^A}{2} H )  
\nonumber \\
& &
+\frac{\kappa}{2\Lambda^2} (\bar{\psi}_L\gamma_\mu\psi_L)
( H^\dagger i\stackrel{\leftrightarrow}{\partial^\mu} H ) 
- \frac{\kappa}{\Lambda^2} (\bar{t}_{R}\gamma_\mu t_R) 
( H^\dagger i\stackrel{\leftrightarrow}{\partial^\mu} H )  
\eea
where we will relate the coupling constant, 
$\kappa$, to $M_H$, $\Lambda$ and $g$ below.

We thus obtain the effective lagrangian, 
\bea
\label{full0}
{\cal{L}}_H & = &  \bar{\psi}_L i\slash{\partial}\psi_L + i\bar{t}_R \slash{\partial} t_R +\partial H^\dagger \partial H 
\nonumber \\
& & + g(\bar{\psi}_L t_R H + h.c.)
- M_H^2 H^\dagger H - \frac{\lambda}{2} (H^\dagger H)^2 
\nonumber \\
& & +\frac{\kappa}{\Lambda^2}(\bar{\psi}_L t_R\bar{t}_{R}\psi_L )
+\frac{2\kappa}{\Lambda^2}(\bar{\psi}_L\gamma_\mu\frac{\tau^A}{2}\psi_L )
(H^\dagger i\stackrel{\leftrightarrow}{\partial^\mu} \frac{\tau^A}{2} H )  
\nonumber \\
& &
+\frac{\kappa}{2\Lambda^2} (\bar{\psi}_L\gamma_\mu\psi_L)
( H^\dagger i\stackrel{\leftrightarrow}{\partial^\mu} H ) 
- \frac{\kappa}{\Lambda^2} (\bar{t}_{R}\gamma_\mu t_R) 
( H^\dagger i\stackrel{\leftrightarrow}{\partial^\mu} H )  
\eea
Performing the super-dilatation
transformation of eq.(\ref{trans0}) we now demand that:
\beq
\delta {\cal{L}}_H  = {\cal{O}}\left(\frac{1}{\Lambda^4} \right)
\eeq
The generated ${\cal{O}}(1/\Lambda^4) $ terms
can be compensated by adding additional $1/\Lambda^4$ terms to
the lagrangian. By continued application of eq.(\ref{trans0}) we 
would generate a power series  of contact interactions 
that are scaled by $\sim 1/\Lambda^{2n}$.  

 First we see that the transformation of the Higgs mass term of eq.(\ref{full0}), 
from eqs.(\ref{trans2}--\ref{transYuk}),   cancels against the first term of the transformed Higgs-Yukawa interaction, provided:
\beq
\label{21}
 g\eta  = \frac{M_H^2 }{\Lambda^2} 
\eeq
(beware: $M_H^2$ is the
negative Lagrangian (mass)$^2$, while $m_h^2 = -2M_H^2$ is the physical 
positive Higgs boson (mass)$^2$ in the broken phase,
and we take the normalization $v^2 = \VEV{H^\dagger H} =(175\; GeV)^2$ )
This establishes
the normalization factor, $\eta$.
It also establishes the relative sign (we assume $\Lambda^2$ positive). 
If we're in the symmetric (broken) phase, 
$M_H^2$ positive (negative), then we have $g\eta >0$ ( $g\eta <0$).
We have the freedom of choosing arbitrary $\eta$ 
since the defining kinetic term invariance involves only $\epsilon$.

One might think we can now take $\Lambda^2$ to be arbitrarily large compared
to $M_H^2$ by adjusting $|\eta| << 1$. However,  the
second term of eq.(\ref{transYuk})
must also cancel against the transformation of the first $\kappa$ term appearing in eq.(\ref{full0}). This requires that:
\beq
\label{above}
\eta{\kappa}= - g,
\qquad \makebox{or, using eq.(\ref{21}):} \qquad \frac{\kappa}{\Lambda^2} = -\frac{g^2}{M_H^2}
\eeq
This is to us a striking result. In the $d=6$ operators
we have a Nambu-Jona-Lasionio component. The Nambu-Jona-Lasinio attractive interaction corresponds to $\kappa >0$ , and we see
in eq.(\ref{above}) that the super-dilatation is then 
consistent only if $M_H^2 < 0$.  
Moreover, to make $\eta$ small requires  taking $\kappa$ large
(or explicitly breaking the super-dilatation symmetry, which we consider in the weak coupling case IV.C).

Finally, the most
interesting relationship, which is the analogue of
the Goldberger-Treiman relationship in a chiral lagrangian (Appendix A),
arises
from the cancellation of the $\sim \epsilon(\bar{\psi}_{L}\theta_R + \bar{\theta}_{L}t_R)\cdot H H^\dagger H$ terms of eqs.(\ref{transquartic}) and (\ref{transYuk})
under the super-dilatation symmetry. This transformation does
not involve $\eta$ and requires:
\beq
\label{cc0}
0 = (\lambda - \half g^2)\frac{\epsilon}{\Lambda^2}(\bar{\psi}_{L}\theta_R + \bar{\theta}_{L}t_R)\cdot H H^\dagger H  + h.c
\eeq
or,
\beq
\label{cc}
\lambda = \half g^2 
\eeq
We emphasize that the critical aspect of our construction is that
the operator shift of $\delta H$ in the quartic Higgs interaction
is cancelling against the super-rotation
(\ie,the ``twist'') of $\delta \psi$ in the Higgs-Yukawa interaction.
Moreover, the pure fermionic shift in  
$\delta \psi \sim \eta\epsilon \theta $, in the Higgs-Yukawa interaction,
\ie, proportional to $\eta$,
cancels against the $\delta H$ shift in the Higgs mass term.
This ties the transformations together into a single structure.
We'll see, in the next section
that this leads to a remarkable effect in the current structure.
The Nambu-Goldstone theorem for the
fermionic shift $\delta \psi \sim \eta\epsilon \theta $, which would naively imply a massless top quark (a ``Goldstino''),
is evaded in the broken phase.

Note that the $\lambda = g^2/2$ relationship refers to the coefficient of the $D=6$ operator,
$(\bar{\theta}_{L}t_R)\cdot H H^\dagger H +h.c. $    We
therefore assume that it applies at the scale $\Lambda$. 
The low energy relationship between the couplings
$g^2$ and $\lambda$ then
depends upon the  renormalization group running from $\Lambda $ to $v_{weak}
\approx 175$ GeV. 
If we ignore the RG running 
then eq.(\ref{cc}) would hold at the weak scale, and 
implies the supersymmetric relationship
 $m^2_{h} = 2\lambda v^2_{weak}  =m^2_{t}$
in the broken phase.  
This is an improvement over the usual NJL result, $m^2_h = 4m^2_t$.

Some comments on
the structure of these higher dimension operators are in order.
Note that we can Fierz rearrange the first term of eq.(\ref{D6}):
\beq
  (\bar{\psi}^a_Lt_{Ra})_i(\bar{t}_{Rb}\psi^b)^i
\rightarrow
- (\bar{\psi}_{iL} \gamma_\mu \frac{\lambda^A}{2} \psi^i_L )
(\bar{t}_{R} \gamma^\mu \frac{\lambda^A}{2} t_{R}) + {\cal{O}(1/N)}
\eeq
where $N=3$ is the number of colors.  
This term is a pure Nambu-Jona-Lasinio interaction as arises in topcolor \cite{BHL,topc} in the form of a (color current)$\times$(color current).
Indeed, massive Yang--Mills boson exchange for a boson of mass $M^2$ and
momentum exchange $q^2 < M^2$ produces the negative sign for (current)$\times$(current) interactions.
A positive sign for the first term 
of eq.(\ref{D6}) is the attractive sign for the Nambu-Jona-Lasinio model, 
and we thus see that the
 attractive sign 
corresponds to the correct (negative) sign for 
topgluon exchange.  However, we see that the isospin 
(current)$\times$(current) interaction (second term of eq.(\ref{D6}))
then has the wrong sign (positive) for a gauge boson exchange.
 We will discuss this ``frustration
of signs'' further in Section IV.A.

Since all of the higher dimension $d=6$ operators  are of the form (current)$\times$(current), they preserve the chirality of the
fermions.  That is, the terms of eq.(\ref{D6}) 
contain
no cross terms of the form $\bar{\psi}_L H t_R(H^\dagger H)^p$.
They thus admit the discrete symmetry, $\psi_L\rightarrow (-1)^N\psi_L$
and $t_R\rightarrow (-1)^{N+1} t_R$.
Operators of mixed chirality can therefore be excluded, or suppressed, on symmetry grounds, though we consider their inclusion 
in Section (IV.C) to improve the prediction for $m_h$.

\section{Current Structure and Nambu-Goldstone Theorem}

The critical aspect of our construction is that
the operator shift of $\delta H$ in the quartic Higgs interaction
is cancelling against the super-rotation
(\ie,  the ``twist'') of $\delta \psi$ in the Higgs-Yukawa interaction.
Moreover, the pure fermionic shift in  
$\delta \psi \sim \eta\epsilon \theta $, in the Higgs-Yukawa interaction,
\ie, proportional to $\eta$,
cancels against the $\delta H$ shift in the Higgs mass term.
This ties the transformations together into a single structure.

The Nambu-Goldstone theorem for a
fermionic shift $\delta \psi \sim \eta\epsilon \theta $, which would naively imply a massless top quark (a ``Goldstino''),
is evaded in the broken phase.
How does our theory evade the existence of a zero-mode
associated with the fermionic shift?  Naively, this would
seem to prohibit a massless top quark. 
In fact, this happens in a subtle way. One must carefully
construct the currents given our use of equations of
motion in $\delta (g\bar{\psi}_L t_R H + h.c.)$.
We therefore wish to clarify the 
the relationship to the
Nambu-Goldstone theorem in the present set up.

We consider, for technical simplicity, a simpler 
``minimal'' transformation defined by $\theta_L=0$:
\bea
\label{trans20a}
\delta \psi^{ia}_L =  - i\frac{\slash{\partial}H^i\theta^a_R}{\Lambda^2}\epsilon ; &&
\qquad  
\delta \bar{\psi}_{L\;ia} = 
i\frac{\bar{\theta}_{Ra}\slash{\partial}H_i^\dagger}{\Lambda^2}\epsilon  ;
\\
\label{trans20b}
\delta t^a_R = \theta^a_R\eta \epsilon ;
&&
\qquad  
\delta \bar{t}_{Ra} = \bar{\theta}_{Ra}\eta \epsilon ;
\\
\label{trans20c}
\delta H^i = \frac{\bar{\theta}_{Ra}\psi^{ia}_L }{\Lambda^2}\epsilon ; 
&&
\qquad  \delta H_i^\dagger = \frac{\bar{\psi}_{Lai}\theta^a_R }{\Lambda^2}\epsilon .
\eea
The parameter $\eta $ is still
fixed by the symmetry
interplay of the Higgs mass term and Yukawa interaction
as in eq.(\ref{21}),
\beq
\label{eta20}
g\eta = \frac{ M_H^2}{\Lambda^2} = -\frac{ \lambda v^2}{\Lambda^2}
\eeq
Consider the top and Higgs
system of the standard model with gauge fields set to zero:
\bea
\label{full30}
{\cal{L}}_H & = &  
{\cal{L}}_{K} + g(\bar{\psi}_L t_R H + h.c.)
- M_H^2 H^\dagger H - \frac{\lambda}{2} (H^\dagger H)
\eea
\beq
\label{kt20}
 {\cal{L}}_{K} = \bar{\psi}_L i\slash{\partial}\psi_L + \bar{t}_R i\slash{\partial} t_R +\partial H^\dagger \partial H 
\eeq
It is readily seen that eqs.(\ref{trans20a}--\ref{trans20c}) is a global invariance of eqs.(\ref{kt20})
up to total derivatives. We presently allow $\epsilon$ to be a
local function of spacetime $\epsilon(x)$ (note that the derivatives in   eq.(\ref{trans20a}) act only upon $H$ and not upon $\epsilon(x)$).
We have:
\bea
\label{transkt20}
\delta (\bar{\psi}_Li\slash{\partial}\psi_L) & = & \frac{(\bar{\psi}_L\theta_R)\cdot \partial^2 H}{\Lambda^2}\epsilon + \frac{(\bar{\psi}_L\gamma_\mu 
\slash{\partial}H \theta_R)  }{\Lambda^2} \partial^\mu\epsilon  +  h.c.+t.d.
\nonumber  \\
\delta (\bar{t}_Ri\slash{\partial}t_R) & = & i(\bar{t}_R\slash{\partial}\theta_R)\eta\epsilon + i(\bar{t}_R\gamma_\mu \theta_R)\eta {\partial}^\mu\epsilon +h.c.+t.d.
\nonumber  \\
\delta (\partial H^\dagger \partial H) & = & -\frac{(\bar{\psi}_L\theta_R)\cdot\partial^2H}{\Lambda^2}\epsilon
+\frac{(\bar{\psi}_L\theta_R )\cdot\partial_\mu H}{\Lambda^2}\partial^\mu\epsilon
+ h.c. + t.d.
\eea
The kinetic terms thus lead to a Noether current:
\beq
J^K_\mu = \frac{ \delta {\cal{L}}_{K} }{ \delta  \partial_\mu \epsilon }
=  
i(\bar{t}_R\gamma_\mu \theta_R)\eta+\frac{(\bar{\psi}_L\gamma_\mu 
\slash{\partial}H \theta_R) }{\Lambda^2}
+\frac{(\bar{\psi}_L\theta_R )}{\Lambda^2}\partial_\mu H + h.c.
\eeq
The symmetry of the full action, as 
we have emphasized, involves a cancellation
of the shift of eqs.(\ref{trans20c})  in the Higgs quartic
interaction against the ``twist'' of eq.(\ref{trans20a}) in the Higgs-Yukawa
interaction.  In calculating the 
transformation of the Higgs-Yukawa interaction, however, 
we make use of  an ``integration by
parts'' and discard total divergences
(and subsequently use the fermion equations of motion).  
This integration by parts in the ``twist'' of eq.(\ref{trans20a})
causes the derivative to act upon the  parameter $\epsilon(x)$:
\bea
\label{trans220}
\delta (-M_H^2 H^\dagger H) & = & 
-\frac{\epsilon}{\Lambda^2} \; M_H^2(\bar{\psi}_{L}\theta_R )\cdot H + h.c
 \\
\delta (-\frac{\lambda}{2}(H^\dagger H)^2) & = & -\frac{\epsilon}{\Lambda^2}\; \lambda(\bar{\psi}_{L}\theta_R )\cdot H H^\dagger H  + h.c.
\\
\label{transYuk20}
\delta (g\bar{\psi}_L t_R H + h.c.) & = & 
g\eta\epsilon(\bar{\psi}_L\theta_R ) H  
+\frac{g^2\epsilon}{2\Lambda^2} (\bar{\theta}_R \psi_L)\cdot \left( H^\dagger H^\dagger H \right) 
 \nonumber \\
& & + \; \frac{g\epsilon}{\Lambda^2}(\bar{\psi}_L t_R)(\bar{\theta}_{R}\psi_L )
-i\frac{g\epsilon}{\Lambda^2}\bar{\theta}_{R}\gamma_\mu t_R 
\left( H^\dagger \stackrel{\leftrightarrow}{\partial^\mu} H \right)  
\nonumber \\
& & 
-i \frac{g}{2\Lambda^2}\bar{\theta}_{R}\gamma_\mu  t_R 
(H^\dagger H)\partial^\mu\epsilon   + h.c. + t.d.
\eea
The last term in eq.(\ref{transYuk20}) shows explicitly that the result of the
integration by parts leads to an additional term $\propto \partial^\mu\epsilon$.
This, in turn, modifies the current, which now becomes:
\beq
J_\mu = \frac{ \delta {\cal{L}}_{H} }{ \delta  \partial_\mu \epsilon }
=   
i(\bar{t}_R\gamma_\mu \theta_R)\left(\eta + \frac{g H^\dagger H}{2\Lambda^2} \right)
+\frac{(\bar{\psi}_L\gamma_\mu 
\slash{\partial}H \theta_R) }{\Lambda^2}
+\frac{(\bar{\psi}_L\theta_R )}{\Lambda^2}\partial_\mu H +h.c.
\eeq
Using the relationship
of  eq.(\ref{eta20}), $g\eta = -\lambda v^2/\Lambda^2$, and the ``Goldberger-Treiman'' analogue,
$\lambda = g^2/2$, the current
can be written:
\beq
\label{current20}
J_\mu{} = \frac{ \delta {\cal{L}}_{H} }{ \delta  \partial_\mu \epsilon }
=   
i\eta(\bar{t}_R\gamma_\mu \theta_R)\left(1 - \frac{H^\dagger H}{v^2} \right)
+\frac{(\bar{\psi}_L\gamma_\mu 
\slash{\partial}H \theta_R) }{\Lambda^2}
+\frac{(\bar{\psi}_L\theta_R )}{\Lambda^2}\partial_\mu H+h.c.
\eeq
The modifcation of the current occurs in the first term
which is associated with the fermionic ``shift''-symmetry of $t_R$.
Again, this arises from the crucial linkage of the
$\delta H$ shift in the quartic Higgs interaction to
the $\delta \psi_L$ shift in the Higgs-Yukawa interaction.

Note that, in the broken phase where $\VEV{H}=v$ the modifcation
of the current has the effect of ``turning off'' the fermionic shift.
Indeed,  we will now see that this has a remarkable effect
in evading the Nambu-Goldstone theorem in the broken phase,
and permitting the top quark to be massive.

Consider the two-point function of our current of eq.(\ref{current20})
with $t_R$:
\beq
S(y) = \int \; d^4x e^{iq\cdot x} \;\partial^\mu \bra{0}T^* J_\mu(x)\;t_R(y)\ket{0}
\eeq
($T^*$ implies anti-commutation in the ordering of fermion fields).
Formally, with $\partial^\mu J_\mu = 0$,
we have, from the $\partial_0$ acting upon the $T^*$ ordering
a $\delta(x^0-y^0)$, and:
\bea
\label{Ward1}
S(y) & = & \int \; d^3x\; e^{iq\cdot x} \bra{0} \{ J_0(x),\;t_R(y)\}\ket{0}
\nonumber \\
& = & \int \; d^3x\; e^{-i\vec{q}\cdot \vec{x}} \bra{0} \{ J_0(\vec{x}),\;t_R(\vec{y})\}_{e.t.}\ket{0}
\nonumber \\
& = &   \bra{0} \{Q,\;t_R(\vec{y})\}\ket{0}
\eea
where the charge operator $Q$ is:
\beq
Q = \int \; d^3x\; J_0(\vec{x}).
\eeq
%where the last line pertains to the $\vec{q}\rightarrow 0$ limit.

In the symmetric phase of the standard model we have
the Higgs VEV,  $\VEV{H} =0$, and we can neglect all
terms in the current that involve $H$.
The charge operator then involves only the first term in $J^K_\mu
= i\eta\bar{t}_R\gamma_\mu \theta_R + h.c.$,
whence it generates a shift in the fermion field:
\beq
\label{Ward2}
\bra{0} \{Q,\;t_R(\vec{y})\}\ket{0} = \eta{\theta_R}
\eeq
On the other hand we have:
\bea
\label{Ward3}
S(y) & = & \int \; d^4x \; e^{iq\cdot x} \;\partial^\mu \bra{0}T^* i\bar{t}_R(x)\gamma_\mu \theta_R\eta\; \; \; t_R(y)\ket{0} \nonumber \\
& = & -\int \; d^4x \; e^{iq\cdot x} \;i\partial^\mu \gamma_\mu S_F(x-y)\theta_R\eta  \nonumber \\
& = & \left. \frac{q^2 + \slash{q}m}{q^2-m^2}\theta_R\eta \right|_{q\rightarrow 0}   
\eea
In the $q^2\rightarrow 0$ limit the consistency 
of eq.(\ref{Ward3}) with eqs.(\ref{Ward1}, \ref{Ward2}) requires that 
the fermion mass satisfy $m=0$.
This is the fermionic  Nambu-Goldstone theorem and it informs us that any 
fermionic action which has a pure fermionic shift symmetry, 
must correspond to a massless fermion.
This is, indeed, the case in the symmetric phase 
in which the top quark is massless and $\VEV{H}=0$.

Naively we might conclude that the top quark is forced by our
symmetry to be a Goldstino
and remain massless, even in the broken phase. However, we have seen 
that the current is modified in a significant way
in the present case:
\beq
J_\mu^{{shift}}  =   
i\eta(\bar{t}_R\gamma_\mu \theta_R)\left(1 - \frac{H^\dagger H}{v^2} \right) +h.c.
\eeq
In the broken phase,  when  $\VEV{H}=v\neq 0$, this implies that the pure fermionic shift operator in the current ``turns off:''
\beq
\label{J2zero}
 J_\mu^{{shift}} = 0 
\eeq
This is a consequence of the interplay between the quartic interaction and
the Higgs-Yukawa interaction in our construction. 
It implies that there can exist dynamical situations in which a Goldstino
is massless in a symmetric phase, but acquires mass in a broken
phase of a theory. The underlying fermionic shift, $\delta \psi =\eta \theta$ is intact, but
the current is modified dynamically to evade the naive Nambu-Goldstone
theorem.  One might wonder what happens for $\VEV{H^\dagger H} \neq v^2$ and $\neq 0$?
This is, of course, and unstable vacuum and the S-matrix derivation fails. 
Of course, the above current algebra analysis serves only as a 
consistency check on our original Lagrangian analysis, which showed
that a vacuum with massive top and Higgs, with $m_t = m_h$, exists.

Indeed, the symmetry yields
$\lambda = g^2/2$ in both phases of the Standard
Model.  However, the fermionic shift part of the current
is nontrivially modified by the quartic-Yukawa interplay
and is $\propto (1-H^\dagger H/v^2)$. It thus vanishes in
the broken phase with $\VEV{H}=v$.   Therefore,
the top quark becomes massive in the broken phase in the usual way, with
the relationship $m_h = m_t$. This is consistent with the Nambu-Goldstone theorem that would
otherwise naively force the top quark to be 
a massless Goldstino.   
This relationship $\lambda = g^2/2$ 
is the analogue of a Goldberger-Treiman'' relationship.
It holds at a high scale, $\Lambda$, and
is subject to renormalization group and higher dimension operator 
effects that can bring the physical 
masses into concordance with $m_h^2 \approx m_t^2/2$.

\vskip 0.25in
\noindent
{\bf Super-Dilatation in the Broken Phase}
\vskip 0.25in

The result or eq.(\ref{J2zero}) helps us to understand 
the symmetry when written in the broken
phase of the theory. We will presently treat
the broken phase in a vectorlike
case, defining $\theta_L=\theta_R$, and we'll ignore the Nambu-Goldstone
bosons of the Higgs field. With the Higgs potential
defined as $V(H) = (\lambda/2)(H^\dagger H - v^2)^2$ we write:
\beq
 H = \left( \begin{array}{c} v + h/\sqrt{2}\\ 0 \end{array} 
 \right)
 \eeq 
The lagrangian then takes the form:
\bea
\label{full40}
{\cal{L}}_H & = &  
{\cal{L}}_{K} + m_t\bar{t}t+ \frac{g}{\sqrt{2}}\bar{t} t h 
- \lambda v^2 h^2 - \frac{\lambda}{\sqrt{2}} h^3 - \frac{\lambda}{8} h^4 
+{\cal{O}}\left(\frac{1}{\Lambda^2}\right)
\eea
and the physical Higgs boson mass is $m_h^2 = 2v^2\lambda$.
The kinetic terms become: 
\beq
\label{kt40}
 {\cal{L}}_{K} = i\bar{t}\slash{\partial}t + \half \partial_\mu h \partial^\mu h 
\eeq
As noted above, the fermionic shift
``shuts off'' in the broken phase.
We assume the symmetry of the broken
phase effective lagrangian 
now involves only the (``twist'') super-transformation 
on $t$ and an operator shift of $h$:
\bea
\label{trans30a}
\delta t & = &  - i\frac{\slash{\partial}h\theta}{\sqrt{2}\Lambda^2}\epsilon ;
\qquad  
\delta \bar{t} = 
i\frac{\bar{\theta}\slash{\partial}h }{\sqrt{2}\Lambda^2}\epsilon  ;
\\
\label{trans30b}
\delta h & = & \frac{\bar{\theta}t }{\sqrt{2} \Lambda^2}\epsilon + h.c. 
\eea
where $\theta$ is a fixed orientation carrying color and spin.
This is readily seen to be a symmetry, modulo
total divergences, of the kinetic terms:
\bea
\label{ktrans30a}
\delta \bar{t}i\slash{\partial}t & = &  \bar{t} \theta\frac{{\partial^2}h}{\sqrt{2}\Lambda^2}\epsilon + h.c. + t.d.
\\
\label{ktrans30b}
\half  \delta \partial_\mu h \partial^\mu h  & = & -\bar{t} \theta\frac{{\partial^2}h}{\sqrt{2}\Lambda^2}\epsilon + h.c. + t.d.
\eea
No use of equations of motion is involved in the invariance
of the kinetic terms. 
The 
transformations of eqs.(\ref{trans30a},\ref{trans30b}) 
also define a symmetry of the full theory: 
\bea
\label{trans230}
\delta (-\lambda v^2 h^2) & = & 
-\sqrt{2} \lambda v^2 h \left(\frac{\bar{\theta}t }{\Lambda^2}\epsilon + h.c.\right)
 \\
 \delta (-\frac{\lambda v}{\sqrt{2}}h^3) & = & -\frac{3\lambda}{{2}}h^2v \left(\frac{\bar{\theta}t }{\Lambda^2}\epsilon + h.c.\right)
\\
 \delta (-\frac{\lambda}{8}h^4) & = & -\frac{\lambda}{2\sqrt{2}}h^3 \left(\frac{\bar{\theta}t }{\Lambda^2}\epsilon + h.c.\right)
\eea
and:
\bea
\label{mtt}
\delta (m_t\bar{t}t) & = & \left( \frac{g^2v^2h}{\sqrt{2}} +  \frac{g^2vh^2}{{2}}\right)     \left(\frac{\bar{\theta}t }{\Lambda^2}\epsilon + h.c.\right)
\\
\label{transYuk40}
\delta (\frac{g}{\sqrt{2}} \bar{t} t h ) & = & 
\left( \frac{g^2vh^2}{{4}} + \frac{g^2h^3}{4\sqrt{2}} \right)    
\left(\frac{\bar{\theta}t }{\Lambda^2}\epsilon + h.c.\right)
\nonumber \\
& & + \frac{g\bar{t}t}{2} \left(\frac{\bar{\theta}t }{\Lambda^2}\epsilon + h.c.\right)
\eea
where in eqs.(\ref{mtt} -- \ref{transYuk40}) 
we have used $m_t=gv$ and the fermionic
equation of motion:
\beq
i\slash{\partial}t +m_tt +\frac{g}{\sqrt{2}} ht = 0
\qquad 
-i\partial^\mu\bar{t}\gamma_\mu + m_t\bar{t} +\frac{g}{\sqrt{2}} h\bar{t} = 0
\eeq
One can readily check, using $\lambda=g^2/2$, 
that eq.(\ref{full40}) is invariant
under eqs.(\ref{trans30a}--\ref{transYuk40}). This
implies $m_t=m_h$.
The four-fermion term in eq.(\ref{mtt}) requires
the addition of the $(\bar{t}t)^2/\Lambda^2$
operator into the action, which in turn modifies the fermion equation of motion,
and will also be invariant under eqs.(\ref{trans30a} -- \ref{trans30b}).
Note that we are always demanding an exact symmetry of kinetic terms,
without use of equations of motion. 
We use only the fermion equation of motion in 
the interaction terms which is equivalent to setting null operators
in the action to zero. The resulting symmetry has the nontrivial
consequence that $\lambda = g^2/2$, equivalently $m_t=m_h$.

\section{Physical Interpretation}

\subsection{Minimal Symmetry and Correct-Sign (current)$\times$(current)
Interactions}

As we noted in section II, the full set of $\kappa$ terms of eq.(\ref{full0}) have an inconsistency if we want to interpret them all in terms of gauge boson mediated (current)$\times$(current) interactions. In particular, the first two terms conflict if we Fierz the first $\kappa$ term into
a (color current)$\times$(color current) interaction, and compare 
to the second 
(isospin current)$\times$(isospin current) interaction. We see that the interactions
have opposite signs; hence we can choose a given sign for $\kappa$ and
one of the two (current)$\times$(current) will have the wrong-sign (positive)
for a Yang-Mills gauge boson mediated
interaction term (negative; note that the ($U(1)$ current)$\times$($U(1)$ current) terms can have either sign, depending upon the $U(1)$ charge assignments of the fields).

However, there is a simple remedy: use the minimal symmetry. The (isospin current)$\times$(isospin current) interaction arises from the presence of $\theta_L$ in our transformation eq.(\ref{transYuk}), while the NJL term arises from both 
$\theta_L$ and $\theta_R$.  We can thus use ``minimal symmetry'' that
contains only the iso-singlet $\theta_R$:
\bea
\label{trans10}
\delta \psi^{ia}_L =  - i\frac{\slash{\partial}H^i\theta^a_R}{\Lambda^2}\epsilon ; &&
\qquad  
\delta \bar{\psi}_{L\;ia} = 
i\frac{\bar{\theta}_{Ra}\slash{\partial}H_i^\dagger}{\Lambda^2}\epsilon  ;
\nonumber  \\
\delta t^a_R = \theta^a_R\eta\epsilon ;
&&
\qquad  
\delta \bar{t}_{Ra} = \bar{\theta}_{Ra}\eta\epsilon ;
\nonumber  \\
\delta H^i = \frac{\bar{\theta}_{Ra}\psi^{ia}_L }{\Lambda^2}\epsilon ; 
&&
\qquad  \delta H_i^\dagger = \frac{\bar{\psi}_{Lai}\theta^a_R }{\Lambda^2}\epsilon .
\eea
The effective lagranian we now obtain is simpler:
\bea
\label{full10}
{\cal{L}}_H & = &  \bar{\psi}_L i\slash{\partial}\psi_L + i\bar{t}_R \slash{\partial} t_R +\partial H^\dagger \partial H 
\nonumber \\
& & + g(\bar{\psi}_L t_R H + h.c.)
- M_H^2 H^\dagger H - \frac{\lambda}{2} (H^\dagger H)^2 
\nonumber \\
& & +\frac{\kappa}{\Lambda^2}(\bar{\psi}_L t_R\bar{t}_{R}\psi_L )
- \frac{\kappa}{\Lambda^2} (\bar{t}_{R}\gamma_\mu t_R) 
( H^\dagger i\stackrel{\leftrightarrow}{\partial^\mu} H )  
\eea
The frustrated (current)$\times$(current) signs are now absent
given that we have banished the isospin interaction. 
We retain the relationships of eqs.(\ref{21},\ref{above},\ref{cc}).
The minimal symmetry may be a more natural framework, and
is also suggested by the weakly interacting case discussed in
Section IV.D below.

\subsection{Strong Coupling}

Let us consider a super-dilatation symmetry in the context of
a UV completion model involving a presumed new strong dynamics at the
(multi-) TeV scale. We assume the broken phase and hence $M_H^2 < 0$,
and $\kappa >0$ and large. 

The largest physically acceptable value we might consider for 
the coupling constant $\kappa$ is of order a critical value
 $\kappa < \kappa_c \sim (4\pi)^2$. Thus, we obtain (recall the physical Higgs boson mass, $125 \; GeV = \sqrt{2}M_H$):
\beq
\Lambda^2 < \kappa_c |M_H^2|/g^2 \qquad \Lambda \lta 4\pi |M_H| \sim 1\; TeV\; \makebox{for} \;g\sim 1.
\eeq
(RG effects allow $g \lta 1$ and $\Lambda$ could thus be upwards of $\sim $ a few TeV within our approximations).

We assume  a strong interaction at the high scale $\Lambda$, containing new physics 
described by the $\kappa$ terms in eq.(\ref{full0}). All of these terms
are current-current interactions.  
We interpret this as a compositeness scale
for the Higgs boson, and the low energy theory is then the standard model.
This differs from the usual NJL model, where $g$ and $\lambda$ diverge
at the compositeness scale, which results in $m_h = 2 m_t$ \cite{BHL},
and typically very large $g$ and $\lambda$ near $\Lambda$.
$\lambda$ is now locked to $g^2$ at the scale $\Lambda$
by our symmetry.

If we also included the normal scalar dilation transformation of the Higgs, eq.(\ref{five20}), then
$\lambda$ is an explicit symmetry breaking scale, and can be arguably small,
\ie, the dilation symmetry becomes a custodial symmetry for small $\lambda$.   
Our basic hypothesis is that the 
super-dilatation is then a surviving  spectrum
symmetry of the solution to the theory. We've reverse engineered
the strong UV theory as a power series in higher dimension operators
from the symmetry.  We have not explicitly shown that such a UV theory
has, in fact, such a solution. 

We can estimate the radiative effects on the Higgs mass  using
 the renormalization group (RG) equations for $\lambda$ and $g$. We include the QCD effects,
 and integrate in the approximation of constant {\em rh} sides
of the RG equations. This yields the leading log effect:
\bea
16\pi^2 \frac{ \partial \lambda}{ \partial \ln(\mu)} & = &
12\lambda^2 + 4N_c\lambda g^2 - 4N_c g^4 \approx (3-2N_c)g^4 \approx -3,
\nonumber \\
16\pi^2 \frac{  \partial g^2 }{\partial \ln(\mu)} & = & \left( 2N_c + 3\right)g^4 - 2(N_c^2-1)g^2g_{QCD}^2
\approx 9g^4-16\times (4\pi \alpha_{QCD})g^2 \approx -13
\nonumber \\
\eea
In the last expressions we've substituted
$\lambda = g^2/2$, $\alpha_{QCD} = 0.11 $ 
and $g= 1$ .  This yields, upon imposing the boundary condition
$\lambda(\Lambda) - g^2(\Lambda)/2 = 0$:
\beq
\label{27}
 \lambda(v_{weak}) - \half g^2(v_{weak})  \approx -\frac{3.5}{16\pi^2}\ln \left(\frac{\Lambda}{v_{weak}}  \right) \approx
-0.04
\eeq
This implies $m_h =\sqrt{2\lambda} v_{weak}\approx 167$ GeV.
This is an improvement over Nambu-Jona-Lasinio inspired models, $\sim 260$ GeV
\cite{BHL}, but 
still too high. The radiative corrections slightly nudge the mass relationship
in the right direction, $m_h < m_t$. Naively, 
with $\Lambda \sim 10^7$ GeV we bring
$m_h\sim 125$ GeV.

\subsection{Chiral Higher Dimension Operators}

We consider presently
modifications to eq.(\ref{full0}) consisting of towers of higher dimension chiral
operators:
\bea
\label{hd}
{\cal{L}}_H & = & {\cal{L}}_{KT}
+ g(\bar{\psi}_L t_R H + h.c.)P(H^\dagger H)
+ M_H^2 H^\dagger H - \frac{\lambda}{2} Q(H^\dagger H)
\eea
where we work in the broken phase and:
\beq
P(H^\dagger H) = \sum_{n=0} c_n \left( \frac{H^\dagger H}{\Lambda^2} \right)^n
\qquad\qquad c_0 = 1
\eeq
\beq
Q(H^\dagger H) = \sum_{n=0} d_n (H^\dagger H)^2 \left( \frac{H^\dagger H}{\Lambda^2} \right)^n
\qquad\qquad d_0 = 1
\eeq
The $P$ operators break the discrete chiral symmetry, $\psi_L\rightarrow (-1)^N\psi_L$
and $t_R\rightarrow (-1)^{N+1}t_R$.
These operators are connected by the super-dilatation to the $Q$'s
as we'll see below. Therefore, we would expect the coefficients
of these to be naturally small.

The Higgs potential minimum is now modified by $Q$,
given by:
\beq
\label{31}
0 
= - M_H^2 + \left. \frac{\lambda}{2} \frac{\partial Q(v)}{\partial (v)^2}\right|_{v=v_{weak}}
\eeq
where $H^\dagger H \equiv v^2$.
At the potential minimum, the top quark mass is:
\beq
m_t = gv_{weak}P(v_{weak}^2)
\eeq
$M_H^2$ is, of course, the curvature 
of the potential at the origin, $v\approx 0$, but the physical Higgs boson mass is determined by the 
curvature at
the potential minimum, $v=v_{weak}$:
\beq
\label{33}
m_h^2 = -M_H^2 + \left.  \frac{\lambda}{4} \frac{\partial^2 Q(v)}{(\partial v)^2}\right|_{v=v_[weak}
\eeq
For the quartic potential we have $Q(v) = v^4$, and
the usual results obtain:
\beq
\makebox{Eq.(\ref{31}):} \;\;\;\ 0 = -M_H^2 + \lambda v^2,
\qquad\qquad
\makebox{Eq.(\ref{33}):} \;\;\;\ m_h^2 = -M_H^2 +3 \lambda v^2.
\eeq
This gives the usual relationship, $m_h^2 = 2M_H^2 = 2\lambda v^2$,
however, this is modified by the presence of the additional terms in $Q(v)$.

If we perform the super-dilatation on the Yukawa and potential terms of eq.(\ref{hd}) we see that
the previous relationship of eq.(\ref{cc0}) is now modified:
\beq
\label{cc2}
0 = \frac{\epsilon}{\Lambda^2}(\bar{\psi}_{L}\theta_R + \bar{\theta}_{L}t_R)\cdot H \left( - \half g^2 H^\dagger H (P(H^\dagger H) )^2
+\frac{\lambda}{2}  \frac{\partial Q(H^\dagger H)}{\partial (H^\dagger H) }\right) + h.c.
\eeq
 Thus, to maintain the super-dilatation symmetry
in the full lagrangian for all values of Higgs fields
we must demand the symmetry condition on the operator towers:
\beq
\half g^2 v^2 (P(v^2) )^2  = \frac{\lambda}{2} \frac{\partial Q(v^2)}{\partial v^2}
\eeq
At the potential minimum, $v= v_{weak}$ we also have from eq.(\ref{31}):
\beq
m_t^2 = g^2 v_{weak}^2 (P(v_{weak}^2) )^2 = \left.{\lambda} \frac{\partial Q(v^2)}{\partial v^2}\right|_{v=v_{weak}} = 2M_H^2. 
\eeq
hence, even in the presence of the tower of operators
we get the result $m_t^2 = 2M_H^2$. But we emphasize that this
is the {\em curvature at the origin of the potential}, and does not give
the physical Higgs mass, which is {\em the curvature at the minimum}.
For the quartic potential this condition implies $m_h^2 = 2M_H^2 = m_t^2$, but in general,
$2M_H^2 \neq m_h^2$, by virtue of the higher dimension terms.

To estimate the size of the effect, consider:
\beq
\frac{\lambda}{2}Q(v^2) = \frac{\lambda}{2}\left(v^4 + \sigma \kappa\frac{v^6}{\Lambda^2}\right)
\eeq 
where the higher dimension operator is assumed to have the strong coupling $\kappa$.
The minimum $v=v_{weak}$ and physical Higgs mass now satisfy
from eqs.(\ref{31},\ref{33}):
\beq
M_H^2 = \lambda v_{weak}^2 + \sigma \kappa \lambda \frac{3v_{weak}^4}{2\Lambda^2},
\qquad\qquad
m_h^2 = -M_H^2 + 3\lambda v_{weak}^2 +  \sigma \kappa \lambda  \frac{15 v_{weak}^4}{2\Lambda^2}.
\eeq
Hence:
\beq
m_t^2 = 2M_H^2 = 2\lambda v_{weak}^2 + 3\sigma \kappa \lambda \frac{v_{weak}^4}{\Lambda^2},
\qquad \qquad
m_h^2 = 2\lambda v_{weak}^2 +  6\sigma\kappa \lambda \frac{v_{weak}^4}{\Lambda^2},
\eeq
and we thus predict:
\beq
m_h^2 \approx  m_t^2 + \frac{3}{2}\sigma \kappa m_t^2 \frac{v_{weak}^2}{\Lambda^2}
\eeq
Thus, we determine $\sigma$ by demanding $m_h^2 \approx 0.5 m_t^2$,
\beq
\sigma \approx -\frac{1}{3} \frac{\Lambda^2}{\kappa v_{weak}^2} \approx -6.5/\kappa 
\approx - 0.04 \kappa_c/\kappa
\eeq
This is a relatively small coefficient compared
to ${\cal{O}}(1)\kappa_c$. It's effect is  
helped by the strong coupling constant $\kappa$.

Of course, the Higgs potential must be stabilized, which can be done by inclusion of still higher dimension operators.
This result is a sketch
to demonstrate that an phenomenologically acceptable
Higgs mass can emerge from the present scheme. The main
strategy is to reduce the curvature of the potential
at the minimum relative to that at the origin.

\subsection{Weakly Coupled Theory}

We can also raise $\Lambda$ and allow $\kappa \sim g^2$ to be perturbatively small. 
Eq.(\ref{above}) then requires $\Lambda \sim M_H$, and thus the Higgs mass floats up to
a large scale.  

To have a small Higgs boson mass term, we can then break the super-dilatation invariance by adding  an explicit  symmetry breaking Higgs mass term. 
This leads to a situation similar to the fine-tuning of the mass
of the lightest Higgs boson in MSSM.  We imagine
$\kappa \sim g^2 \sim 1$ and choose $\Lambda \sim M_H$ large. We then add a symmetry breaking
term of order $(-M_H^2 + \delta M_H^2)H^\dagger H$ to the lagrangian.  Thus we
are fine-tuning a small Higgs boson mass 
to a level of $\sim \delta M_H^2/M_H^2$ by demanding a cancellation between
a large Higgs boson mass that respects the symmetry and another that breaks it.
 If we take $\Lambda \sim 10^4$ TeV, then our above RG estimate of
eq.(\ref{27}) yields the phenomenologically acceptable result 
$m_h^2 \approx m_t^2/2$.  
 
 This is more naturally framed in the minimal version of the symmetry,
eq.(\ref{trans10}). We are essentially violating the $\delta t_R$ transformation
(corresponding to $\eta\epsilon$ while retaining the $\epsilon$ transformation.
That is, the ``most minimal'' transformation is:
\bea
\delta \psi^{ia}_L =  - i\frac{\slash{\partial}H^i\theta^a_R}{\Lambda^2}\epsilon ; &&
\qquad  
\delta \bar{\psi}_{L\;ia} = 
i\frac{\bar{\theta}_{Ra}\slash{\partial}H_i^\dagger}{\Lambda^2}\epsilon  ;
\nonumber  \\
\delta H^i = \frac{\bar{\theta}_{Ra}\psi^{ia}_L }{\Lambda^2}\epsilon ; 
&&
\qquad  \delta H_i^\dagger = \frac{\bar{\psi}_{Lai}\theta^a_R }{\Lambda^2}\epsilon .
\eea
The effective lagrangian we now obtain is:
\bea
{\cal{L}}_H & = &  \bar{\psi}_L i\slash{\partial}\psi_L + i\bar{t}_R \slash{\partial} t_R +\partial H^\dagger \partial H 
\nonumber \\
& & + g(\bar{\psi}_L t_R H + h.c.)
- M_H^2 H^\dagger H - \frac{\lambda}{2} (H^\dagger H)^2 
\nonumber \\
& & +\frac{\kappa}{\Lambda^2}(\bar{\psi}_L t_R\bar{t}_{R}\psi_L )
- \frac{\kappa}{\Lambda^2} (\bar{t}_{R}\gamma_\mu t_R) 
( H^\dagger i\stackrel{\leftrightarrow}{\partial^\mu} H )  
\eea
The symmetry is broken, but
we retain only the most interesting relationship eq.(\ref{cc}).

\newpage

\section{Conclusions}

In summary, we've shown that the following lagrangian is invariant under the transformation
of eq.(\ref{trans0}) up to total divergences to order $1/\Lambda^4$:
\bea
{\cal{L}}_H & = &  \bar{\psi}_L i\slash{\partial}\psi_L + i\bar{t}_R \slash{\partial} t_R +\partial H^\dagger \partial H 
\nonumber \\
& & + g(\bar{\psi}_L t_R H + h.c.)
- M_H^2 H^\dagger H - \frac{\lambda}{2} (H^\dagger H)^2 
\nonumber \\
& & +\frac{\kappa}{\Lambda^2}(\bar{\psi}_L t_R\bar{t}_{R}\psi_L )
+\frac{2\kappa}{\Lambda^2}(\bar{\psi}_L\gamma_\mu\frac{\tau^A}{2}\psi_L )
(H^\dagger i\stackrel{\leftrightarrow}{\partial^\mu} \frac{\tau^A}{2} H )  
\nonumber \\
& &
+\frac{\kappa}{2\Lambda^2} (\bar{\psi}_L\gamma_\mu\psi_L)
( H^\dagger i\stackrel{\leftrightarrow}{\partial^\mu} H ) 
- \frac{\kappa}{\Lambda^2} (\bar{t}_{R}\gamma_\mu t_R) 
( H^\dagger i\stackrel{\leftrightarrow}{\partial^\mu} H )  
\eea
where:
\beq
\lambda = \half g^2 \qquad \makebox{ and} \qquad   g^2\Lambda^2 = -\kappa M_H^2 .
\eeq
The super-dilatation symmetry is a bashing together of
supersymmetry and dilatation or reparameterization invariance. It involves a super-rotation, or ``twist,'' amongst
the Higgs and top quark fields, essentially forcing the Higgs to become a superpartner $\psi_L=(t,b)_L$.  The leading $D=4$ terms include the invariant gaugeless kinetic terms of Higgs, and top doublet and singlet fields, and the Higgs potential and Higgs-Yukawa interaction. 

Our key result is that the top Yukawa coupling constant can be directly related to the Higgs quartic coupling constant by the super-dilatation 
symmetry, yielding the unrenormalized relationship 
in the broken phase $m_h = m_t$ (note $m_h^2 = 2|M_H^2|$).  We've taken pains to
show the consistency of this and the symmetry
with the Nambu-Goldstone theorem for fermions. 
We have also explicitly demonstrated the symmetry in the broken phase
where the fermionic shift ``turns off.''  The broken phase is a simpler
case to analyze. Though equations of motion are deployed in interaction
terms, this is a bona fide symmetry as it implies the nontrivial relationship 
$\lambda = g^2/2$ or $m_t=m_h$.

With
radiative and/or small higher dimension operator effects, we can 
in principle obtain the observed $m^2_h \approx 0.5 m^2_t$. 
However, we feel that exploring generalizations of the standard model
with this kind of symmetry may lead to a tree level relationship,
 $m_h^2 = m_t^2/2$ .
A remarkable aspect of this is that the super-dilatation symmetry  is already seen to be present in
the existing (gaugeless) kinetic terms of the top-Higgs system.
This may be a hint of the onset of a larger extended supersymmetry
in a non-standard form. 

The super-dilatation symmetry seems also to reflect a Nambu-Jona-Lasinio-like structure at $\Lambda$ which may be a natural harbinger of a composite Higgs boson, composed of $t\bar{t}$, but with a new dynamics, different
than the conventional NJL setting \cite{BHL}.

\vskip .1in
\section{Acknowledgements}
\vskip .1in
\noindent
I thank Bill Bardeen, Roni Harnik, Adam Martin, Graham Ross and Geban Stavenga for discussions.

\vskip .5in

\newpage
\noindent
{\bf Appendix A:  Dilatonic Higgs} 
\vskip 0.25 in

An idea that is currently receiving revived attention 
is that the Higgs boson is an approximate dilaton,
or ``pseudo-dilaton'' (see, \eg, \cite{Kitano} and references therein). 
There is, of course, a fundamental distinction between a ``scale invariant Higgs field''
and a ``dilatonic Higgs field.'' 
A scale invariant Higgs field has a vanishing mass term, but can
have a nonvanishing gauge, quartic and Yukawa couplings.
To qualify as a (pseudo) dilatonic Higgs boson, the Higgs potential must be (approximately) flat. 

Consider the pure Higgs lagrangian (no gauge fields or 
couplings to fermions):
\beq
{\cal{L}}_0 =  \partial_\mu H^\dagger \partial^\mu H  - \frac{\lambda}{2}(H^\dagger H- v^2)^2
\eeq
As usual, the groundstate has a minimum for:
\beq
\VEV{H^i} = \theta^i \qquad \makebox{where} \qquad \theta^i = (v, 0)
\eeq
(of course, $\theta^i$ can be rotated  under $SU(2)_L\times U(1)$
and has an arbitrary orientiation, but $\theta^\dagger \theta = v^2$). 
We can now take the limit 
$\lambda \rightarrow 0$,  corresponding to a flat Higgs potential.
The Higgs potential has served the role of 
an ``applied external magnetic field to a spin system,'' and aligns the spins.
In the present case it pulls $H^i$ to the minimum VEV. But once we take
the limit $\lambda \rightarrow 0$ the lagrangian acquires a ``shift symmetry,"
\beq
\label{five10}
\delta H^i = \theta^i \epsilon \qquad \longrightarrow \qquad \delta \; \partial_\mu H^\dagger \partial^\mu H = 0
\eeq
Note that the alignment $\theta^i$ is held fixed (and remains arbitrary)
and the shift is parameterized by the variable $\epsilon$.
The Noether current is:
\beq
J_\mu = \frac{\delta {\cal{L}}_0}{\delta \partial^\mu \epsilon}
= \theta^\dagger \partial_\mu H + H^\dagger \partial_\mu \theta
\eeq
We see that $\theta$ is a defining part of the current. 
If we view $\theta$ as co-rotating with $H$ under the global $SU(2)\times U(1)$
transformations, the charge $\int d^3x J_0$ commutes with the 
gauge group.

In the broken phase of the theory we consider the small fluctuations around the
Higgs VEV, and we can extract the broken phase Higgs field as:
\beq
\frac{1}{2v}(\theta^\dagger H + H^\dagger \theta) = v + \frac{h}{\sqrt{2}}
\eeq
In the broken phase of the theory we thus see that:
\beq
 J_\mu \rightarrow \sqrt{2}v\partial_\mu h
\eeq
This becomes the ``dilatonic current'' of the standard model Higgs boson
in the broken phase.  This is the only possible source
of the dilatonic current, which has a single derivative, as the usual scale
current, $S_\mu = x^\nu T_{mu\nu}$, with improved stress tensor $T_{\mu\nu}$,
is bilinear in derivatives. The Higgs $h$ in the limit $\lambda\rightarrow 0$
is now a Nambu-Goldstone boson 
of spontaneous scale symmetry breaking, inherited from the shift symmetry
of the flat potential, \ie, $h$ has become a ``dilaton.''

The dilatonic nature of the Higgs implies that fields that acquire
masses $\propto v$ are scale invariant in the sense of spontaneous
scale breaking. That is, we can perform a scale transformation
which would shift mass terms, but we can then undo this rescaling by a compensating shift in $h$. To see this, consider the top quark mass term:
\beq
g\bar{\psi_L}{t_R} H +h.c. \;\; \longrightarrow \;\; m_t\bar{t}t\left(1 + \frac{h}{\sqrt{2}v}\right)
\eeq
Under an infinitesimal 
scale transformation we have $t(x) \rightarrow (1-\epsilon)^{3/2}
t(x')$ and $h(x) \rightarrow (1-\epsilon)h(x') $ where $x_\mu=(1+\epsilon)x'{}_\mu$, $d^4x = (1+\epsilon)^4d^4x'$.  Hence the action transforms as:
\bea
S_0 & = & \int d^4x\; m_t\bar{t}t(x) \left(1 + \frac{h(x)}{\sqrt{2}v}\right) \nonumber \\
& \rightarrow & \int d^4x'\left((1+\epsilon) m_t\bar{t}t(x') + m_t\bar{t}t(x')\frac{h(x')}{\sqrt{2}v}\right) 
\eea
The latter expression exhibits the fact that the $d=4$ Higgs-Yukawa
interaction is scale invariant while the $d=3$ mass term 
would scales
as $m\bar{t}t\rightarrow (1-\epsilon)m\bar{t}t $.

However, with the dilatonic shift symmetry
we can compensate the rescaled mass term by a shift in the Higgs-dilaton field.
That is, under:
\beq
h(x')  \rightarrow  h(x') - \sqrt{2}v \epsilon
\eeq
we see that:
\beq
\int d^4x'\left((1+\epsilon) m_t\bar{t}t(x') + m_t\bar{t}t(x')\frac{h(x')}{\sqrt{2}v}\right) 
\rightarrow 
\int d^4x' \left( m_t\bar{t}t(x') + m_t\bar{t}t(x')\frac{h(x')}{\sqrt{2}v}\right)
= S_0
\eeq
Hence, the simultaneous application of
the
conformal transformation and Higgs shift symmetry
allows us to maintain the symmetry
of the top quark mass term; the scale symmetry is
broken spontaneously with the Higgs boson playing
the role of the Nambu-Goldstone mode. The same 
invariance applies to the masses of all fermions, and of the gauge fields, $W$ and $Z$.  The Higgs self-interactions that
involve nonzero $\lambda$ would not be invariant under scale transformations
with dilatonic shifts in h. The shift symmetry is what generates the
Nambu-Goldstone pole.

A consequence of the approximate dilatonic
nature of the Higgs are low energy theorems
for Higgs boson couplings to gauge and fermion fields.
The effective lagrangian for the masses of the $W$--boson, $Z$--boson
arises from the Higgs kinetic term, 
\bea
{\cal{L}} & = & (D_\mu H)^\dagger(D^\mu H)\;\; \longrightarrow\;\;
  M_W^2 {W}_\mu^+ {W}_\mu^- + \frac{1}{2}M_Z^2 {Z}_\mu{Z}^\mu
\eea
where the Standard Model coupling
constants, $g_1$ and $g_2$, imply:
\beq
M_W^2 =  \half g_2^2 v^2   \qquad \qquad M_Z^2 =  \frac{1}{2}(g_1^2 + g_2^2) v^2
\eeq
The Higgs tree-level couplings are obtained simply
by shifting $v$ by the dynamical Higgs field, $h$:
\beq
v \rightarrow v +  \frac{h}{\sqrt{2}}
\eeq
to obtain:
\bea
\delta {\cal{L}} 
&= & \frac{h}{\sqrt{2} v} \left( 2M_W^2   {W}_\mu^+ {W}_\mu^-
+ M_Z^2  {Z}_\mu {Z}^\mu \right)
\eea
These are the leading tree-level couplings of the Higgs field to the
gauge bosons and are measured at the LHC in concordance with
the prediction of a standard model Higgs boson.

The scale anomalies (running coupling constants) break the
scale symmetry. However, this leads to a collection of low energy
theorems for the Higgs coupled to gauge bosons, such as $h\rightarrow gg$
and $h\rightarrow \gamma\gamma;\; \gamma Z $, etc.

Note that the $\lambda \rightarrow 0$ limit
 is the opposite of a nonlinear $\sigma$-model limit in which $\lambda\rightarrow \infty$ as $v$ is held fixed. This latter case leads 
 to the Nambu-Goldstone bosons of electroweak symmetry breaking, but an infinitely massive Higgs boson.
In the case of a dilatonic Higgs, we package both the standard model $SU(2)\times U(1)$ symmetry together with the scale invariance and both are spontaneously broken. 

The Higgs boson is then a (pseudo) dilaton if
the shift transformation is a (approximate) symmetry of the action.
Fundamentally it stems
from the exact shift or modular symmetry of the gaugeless Higgs kinetic term:
\beq
\label{five20}
\delta H^i = \theta^i \epsilon \qquad \longrightarrow \qquad \delta \; \partial_\mu H^\dagger \partial^\mu H = 0
\eeq
A key point we wish
to emphasize here is that $\epsilon$ is the infinitesimal parameter of the transformation, while $\theta^i$ is  held fixed.
The spurion $\theta^i$ defines an arbitrary direction
in isospin space, a ``ray,'' and the shift moves the field along 
this direction in field space.
We take $\theta^i $ to have the same mass dimension
as the Higgs, \ie, dimensions of mass. Here  
$i$ refers to isospin and  $\theta^i$ is a normalized isospinor spurion,
$\theta^\dagger \theta = v^2$, where we, after-the fact, choose the alignment 
$\theta^i = (v, 0)$.     
Eq.(\ref{five20}) is a symmetry
of the gaugeless Higgs boson kinetic terms,  $\partial H^\dagger \partial H$.
In such a theory the shift symmetry is exact.

Our present scheme is essentially a
``super'' generalization of this shift symmetry  of eq.(\ref{five20}),
involving shifts and rotations together with
the top quark doublet and singlet fields. For us, the shift
in the Higgs field becomes an operator, $\delta H \propto
\bar{\psi}_L\theta_R + \bar{\theta}_L t_R$, and fermions
shift as well, $\delta \psi_L \propto \theta_L$ and  $\delta t_R \propto \theta_R$. In addition we have the super-rotation, hence we dub this a
``super-dilataion.''

\newpage
\noindent
{\bf Appendix B: Analogy to a Chiral Lagrangian}
\vskip 0.25 in

To orient the reader to our present strategy we quickly
review a familiar
derivation of a toy $\pi$ chiral lagrangian from the ``bottoms-up.''
Consider the kinetic terms:
\beq
{\cal{L}}_K = \bar{\psi}_L i\slash{\partial} \psi_L +  \bar{\psi}_R i\slash{\partial} \psi_R
+ \half \partial_\mu \pi \partial^\mu \pi
\eeq
We'll consider the RH-chiral symmetry:
\bea
\delta \psi_L & = & 0  
\nonumber  \\
\delta \psi_R & = & i\theta\psi_R   
\qquad  
\delta \bar{\psi}_{R} = -i\theta\bar{\psi}_R
\nonumber  \\
\delta \pi & = & f_\pi \theta
\eea
and we demand the lagrangian is invariant under
this global transformation:
\beq
\delta {\cal{L}}_K  = 0
\eeq
The RH-chiral current is:
\beq
-\frac{\delta {\cal{L}}_K}{\delta \; \partial^\mu \theta} 
= \bar{\psi}_R \gamma_\mu \psi_R -   f_\pi \partial_\mu \pi
\eeq
and we assume 
$f_\pi$ is ``determined from experiment,'' \eg, $\pi \rightarrow \mu \nu$
(of course, in the real world this is the left-handed current).

Consider the interactions consisting of
a massive ``nucleon'' coupled to pion:
\beq
{\cal{L}}_V = M \bar{\psi} \psi -  ig\pi \bar{\psi}\gamma^5 \psi
= M \bar{\psi}_L \psi_R -  ig\pi \bar{\psi}_L \psi_R + h.c.
\eeq
We perform the RH-chiral transformation transformation:
\beq
\delta {\cal{L}}_V = \left(i\theta M  -  igf_\pi \theta  + 
  g\pi f_\pi \theta \right) \bar{\psi}_L \psi_R
+  h.c.
\eeq
so:
\beq
\delta {\cal{L}}_V = 0 \qquad \longrightarrow \qquad  g = \frac{M}{f_\pi}
\eeq
which is the Goldberger-Treiman relation.

However, we still must cancel the
``higher order term'' $ \propto \pi \theta \bar{\psi}_L \psi_R$. 
We thus include an ${\cal{O}}(\pi^2)$ term:
 \beq
{\cal{L}}_V \rightarrow  M \left( 1 - \frac{i\pi}{f_\pi} + c \frac{\pi^2}{f^2_\pi} \right) \bar{\psi}_L \psi_R + h.c.
\eeq
Now:
\beq
\delta {\cal{L}}_V \rightarrow  M\left(i\theta -i\frac{f_\pi\theta  }{f_\pi}
+\frac{\pi\theta  }{f_\pi}
 + 2c \frac{\pi}{f^2_\pi} f_\pi\theta   +    ic \frac{\pi^2}{f^2_\pi} f_\pi\theta
\right)\bar{\psi}_L \psi_R
h.c.
\eeq
so:
\beq
\delta {\cal{L}}_V = 0 \qquad \longrightarrow \qquad  g = \frac{M}{f_\pi}, \qquad c = -\half
\eeq
But, now
we must cancel higher order term $ \propto \pi^2 \theta \bar{\psi}_L \psi_R$
which implies an ${\cal{O}}(\pi^3)$ interaction, and so-forth.

We can sum the resulting power series
and we find, iteratively, the solution:
\beq
{\cal{L}}_V  
= M \bar{\psi}_L U \psi_R + h.c. \qquad \qquad U = \exp(i\pi/f_\pi)
\eeq
whence,
\beq
{\cal{L}} = \bar{\psi}_L i\slash{\partial} \psi_L +  \bar{\psi}_R i\slash{\partial} \psi_R
+ \frac{f_\pi^2}{2} \partial_\mu U^\dagger \partial^\mu U + M \bar{\psi}_L U \psi_R + h.c.
\eeq
and we have obtained the``nonlinear $\sigma$-model lagrangian." 

Our present strategy is similar. We begin with the super-dilatational invariance
of the top-Higgs kinetic terms. We then analyze the transformation of the
Higgs-Yukawa, Higgs mass and quartic interactions. We demand overall
invariance of the lagrangian. We thus find the ``Goldberger-Treiman'' relationship, $\lambda = g^2/2$, which implies $m_t=m_h$
in the broken phase. This induces higher dimension operators. Ultimately,
we expect to sum the tower of operators, though in the present
case we expect that these arise via new dynamics, such as heavy recurrences of
composite Higgs bosons and vector--like top quarks.

\newpage 

%%%%%%%%%%%%%%%%%%%%%%%%%%%%%%%%%%%%%%%%%%%%%%%%%%%%%%%%%%%%%%%%%%%%%%%
%\newpage


\begin{thebibliography}{99}  
  
  
\bibitem{yamawaki} 
  V.~A.~Miransky, M. Tanabashi, and K. Yamawaki,
  Phys. Lett. {\bf B221}, 177 (1989);
  Mod. Phys. Lett. {\bf A4}, 1043 (1989);\\
  W.~J.~Marciano,
  Phys. Rev. Lett. {\bf 62}, 2793 (1989);\\
  Y.~Nambu, Enrico Fermi Institute Report No. 89-08, 1989 
  (unpublished); in {\it Proceedings of the 1989 Workshop 
   on Dynamical Symmetry Breaking}, edited by T. Muta and K. Yamawaki 
  (Nagoya University, Nagoya, Japan, 1990).
  
\bibitem{BHL} 
W.~A.~Bardeen, C.~T.~Hill, and M.~Lindner,
  Phys. Rev. {\bf D41}, 1647 (1990).
\bibitem{topc}
  C.~T.~Hill,
  Phys.\ Lett.\ B {\bf 266}, 419 (1991);
  C.~T.~Hill,
  Phys.\ Lett.\  B {\bf 345}, 483 (1995). 

%\bibitem{Seiberg} 
%  Z.~Komargodski and N.~Seiberg,
%  %``From Linear SUSY to Constrained Superfields,''
%  JHEP {\bf 0909}, 066 (2009)
%  [arXiv:0907.2441 [hep-th]].
%  %%CITATION = ARXIV:0907.2441;%% 

\bibitem{Georgi} 
H. Georgi, Nucl. Phys. {\bf B348}, 293 (1991);
 Phys.  Lett. B {\bf 240}, 447 (1990);\\
 M. Neubert, Phys. Rept. {\bf 245}, 259 (1994)\\
A. V. Manohar and M. B. Wise, Camb. 
Monogr. Part. Phys. Nucl. Phys. Cosmol. {\bf 10}, 1
(2000).

\bibitem{LukeManohar}
M. E. Luke and A. V. Manohar, Phys. Lett. B 286, 348 (1992).

\bibitem{Heinonen} 
  J.~Heinonen, R.~J.~Hill and M.~P.~Solon,
  %``Lorentz invariance in heavy particle effective theories,''
  arXiv:1208.0601 [hep-ph].
  %%CITATION = ARXIV:1208.0601;%%  
  
\bibitem{Kitano} 
  See: S.~Iso and Y.~Orikasa,
  ``TeV Scale B-L model with a flat Higgs potential at the Planck scale - in  view of the hierarchy problem -,''
  arXiv:1210.2848 [hep-ph].\\
  %%CITATION = ARXIV:1210.2848;%%
  T.~Abe, R.~Kitano, Y.~Konishi, K.~-y.~Oda, J.~Sato and S.~Sugiyama,
  ``Minimal Dilaton Model,''
  arXiv:1209.4544 [hep-ph].\\
  %%CITATION = ARXIV:1209.4544;%%
 K.~A.~Meissner and H.~Nicolai,
  %``Renormalization Group and Effective Potential in Classically Conformal Theories,''
  Acta Phys.\ Polon.\ B {\bf 40}, 2737 (2009)
  [arXiv:0809.1338 [hep-th]].
  and references therein.
  %%CITATION = ARXIV:0809.1338;%%
  
\bibitem{Wess} 
  J.~Wess and J.~Bagger,
  %``Supersymmetry and supergravity,''
  Princeton, USA: Univ. Pr. (1992) 259 p;
  see eqs(3.3-3.9).

\end{thebibliography}
\end{document}